\documentclass[letterpaper,journal]{IEEEtran}
\usepackage[T1]{fontenc}
\usepackage[utf8]{inputenc}
\usepackage{amsmath,amsfonts,amssymb,mathtools,amsthm}
\usepackage{array,booktabs,multirow,tabularx,adjustbox}
\usepackage[table]{xcolor}
\usepackage{graphicx}
\usepackage{tikz}
\usetikzlibrary{arrows.meta,positioning,calc}
\usepackage{cite}
\usepackage{url}
\usepackage{stfloats}
\usepackage{microtype,nicefrac,pifont,enumitem}
\usepackage[hidelinks]{hyperref}
\usepackage[capitalize,noabbrev]{cleveref}

\newcolumntype{C}{>{\centering\arraybackslash}X}
\begin{document}
\title{RaGCD: Reliability-Aware 2D-3D Multimodal Generalized Category Discovery}
\author{%
  Junyong Liu\textsuperscript{1}\enspace
  \\[0.2em]
  {\normalfont\textsuperscript{1}Beijing University of Posts and Telecommunications}
}
\maketitle
\raggedbottom

\begin{abstract}
Generalized Category Discovery (GCD) aims to recognize known categories while discovering novel ones from partially labeled data. Existing GCD studies primarily focus on unimodal observations or image--text interactions, leaving the complementarity between visual appearance and explicit 3D geometry underexplored. To exploit complementary visual appearance and spatial geometry, we investigate GCD with paired images and point clouds. However, the relative reliability of these modalities varies across samples, so unreliable evidence can compromise both multimodal fusion and cross-modal knowledge transfer. This challenge is particularly pronounced for novel categories, where labeled supervision is unavailable. To address it, we propose RaGCD, a reliability-aware framework for 2D--3D multimodal GCD. Specifically, Reliability-aware Modality Allocation (RMA) combines prediction-derived reliability cues with modality-specific features to adaptively allocate image and point-cloud contributions at the channel level. Reliability-conditioned Alignment (RCA) further reweights feature-level and prediction-level alignment and selects reliable observations for prototype-level alignment, regulating cross-modal interaction according to sample reliability. Extensive experiments on ModelNet40, Washington RGB-D, and ShapeNet-ViPC demonstrate that RaGCD outperforms the compared methods in overall and novel-category clustering accuracy. Ablation studies support the complementary roles of RMA and RCA, while controlled corruption experiments show that modality allocation shifts toward the less corrupted modality. These results support the effectiveness of reliability-aware interaction for category discovery from paired appearance and geometry.
\end{abstract}


\section{Introduction}
\label{sec:intro}

Generalized Category Discovery (GCD)~\cite{vaze2022generalized} aims to jointly recognize known categories and discover novel categories from partially labeled data. Unlike conventional supervised recognition, GCD assumes that labeled samples cover only a subset of categories, while the unlabeled set contains both known and novel categories. The model must therefore transfer useful knowledge from labeled examples without restricting the unlabeled data to the known-category label space. Its central challenge is to learn a discovery representation that preserves discrimination among known categories while revealing the semantic structure of previously unseen categories. This setting is relevant to open-world object recognition, where models must recognize known categories while discovering previously unseen ones.

Most existing GCD studies~\cite{vaze2022generalized,
wen2023parametric,selex,he2025seal,feng2025generalized,
tang2025dissecting} focus on unimodal observations.
Recent research incorporates complementary linguistic information
through text retrieval and cross-modal learning~\cite{may2023clip,
zheng2024textual,su2025multimodal}, text embedding synthesis and
prompt tuning~\cite{wang2025get,add2}, and structured
vision--language representation learning~\cite{solanki2024graphvl,
ssr2gcd}. Beyond image--text settings,
Cheng et al.~\cite{cheng2026multimodal} investigate generalized
defect category discovery using images and point clouds in
industrial scenarios. These advances motivate further exploration
of appearance--geometry complementarity for object category
discovery. In this work, we focus on how to exploit paired images
and point clouds under partial category supervision while
accounting for their sample-dependent reliability.

Images and point clouds describe different aspects of the same object. Images capture texture, color, and appearance-level semantics, whereas point clouds provide explicit information about shape and spatial configuration. These modalities can help resolve different types of ambiguity. For example, illumination changes and reflections may obscure visual appearance while leaving geometric evidence informative. Conversely, structurally similar objects, such as a closed laptop and a hardcover book, may be difficult to distinguish from sparse point clouds alone, whereas their visual appearance can provide additional semantic cues. Integrating the two modalities can therefore offer a richer basis for organizing object categories than relying on either observation alone. These properties motivate a paired 2D--3D GCD setting that combines appearance and geometry for known-category recognition and novel-category discovery.

Nevertheless, the availability of complementary modalities does not ensure that their information is equally useful for every sample. Images may contain ambiguous or degraded appearance cues, while point clouds may suffer from sparsity, incompleteness, or geometric ambiguity. The relative usefulness of the two modalities can consequently vary across objects and observation conditions. A modality that provides discriminative evidence for one sample may be less informative for another. We refer to this sample-dependent asymmetry as \emph{cross-modal reliability mismatch}. In GCD, this mismatch is particularly challenging for novel categories because labeled supervision is unavailable to directly identify misleading evidence or correct the resulting learning signals.

Cross-modal reliability mismatch affects both the construction of multimodal representations and the interaction between modality branches. For representation construction, fixed fusion assigns the same relative importance to modalities regardless of their sample-specific quality. As a result, ambiguous evidence from one modality may interfere with informative features from the other. Learned fusion can provide greater flexibility, but a feature-dependent weighting mechanism alone does not explicitly describe how predictive uncertainty and cross-modal disagreement should influence modality contributions. This motivates incorporating prediction-derived reliability cues into the allocation of appearance and geometric information.

For cross-modal interaction, alignment encourages the two branches to exchange information and learn compatible representations. However, the usefulness of this interaction also depends on the evidence available for each sample. When one modality provides an unreliable prediction or an ambiguous representation, uniformly encouraging agreement may transfer its errors to the other branch. Moreover, a fused prediction can remain uncertain even after the modality contributions have been adjusted. Improving fusion therefore does not automatically make every cross-modal learning target reliable. This concern is especially relevant to novel categories, whose structure must be learned from unlabeled observations and model-generated training signals. Effective multimodal discovery requires considering both which modality should contribute more to the representation and which samples should contribute more to alignment.

To address these challenges, we propose \textbf{RaGCD}, a reliability-aware framework for 2D--3D multimodal GCD. RaGCD first extracts separate appearance and geometric representations from paired images and point clouds. It then introduces \textbf{Reliability-aware Modality Allocation (RMA)} to adapt their contributions to individual samples. RMA combines modality-specific features with prediction-derived reliability cues, including predictive uncertainty and cross-modal disagreement. A learned channel-wise allocation is combined with an explicit reliability prior to produce the fused discovery representation. This design allows the model to adjust modality contributions using both feature content and predictive evidence, while retaining the flexibility to assign different preferences across feature channels.

RaGCD further introduces \textbf{Reliability-Conditioned Alignment (RCA)} to regulate cross-modal learning at the feature, prediction, and prototype levels. Feature alignment encourages correspondence between paired image and point-cloud embeddings, while prediction alignment transfers fused predictions to the unimodal branches. Prototype alignment additionally promotes consistency between their category-level structures. RCA reweights feature and prediction losses according to sample reliability and selects reliable observations for prototype construction, reducing the relative influence of uncertain or inconsistent samples on these interactions. RMA and RCA thus address complementary aspects of multimodal discovery: RMA determines how modality evidence contributes to the fused representation, while RCA determines how training samples contribute to cross-modal alignment. Their combination incorporates reliability information into both representation construction and knowledge transfer.

We conduct extensive experiments on ModelNet40, Washington RGB-D, and ShapeNet-ViPC, covering synthetic CAD objects, real RGB-D observations, and partial geometry. RaGCD achieves the highest overall and novel-category clustering accuracy among the compared methods on all three datasets. Component ablations examine the individual and combined effects of RMA and RCA, while comparisons with alternative fusion strategies and uniformly weighted alignment assess the role of reliability conditioning. Controlled modality corruption experiments further show that the learned allocation shifts toward the less corrupted modality. Additional analyses examine cross-modal prediction behavior and sensitivity to key model and data settings.

Our main contributions are summarized as follows:
\begin{itemize}
    \item We investigate paired 2D--3D multimodal GCD and address
    sample-dependent cross-modal reliability mismatch, highlighting
    its implications for both modality fusion and cross-modal
    alignment under partial category supervision.

    \item We propose RaGCD, which combines reliability-aware modality
    allocation with reliability-conditioned alignment. RMA adapts
    channel-wise image and point-cloud contributions using
    predictive cues and modality features, while RCA regulates
    feature-, prediction-, and prototype-level interaction
    according to sample reliability.

    \item Experiments on three paired 2D--3D datasets demonstrate
    improvements in overall and novel-category clustering accuracy.
    Component ablations, comparisons with alternative interaction
    strategies, and controlled corruption experiments examine
    the effectiveness and behavior of the proposed framework.
\end{itemize}
\section{Related Work}
\label{sec:related_work}

\subsection{Generalized Category Discovery}

Generalized Category Discovery (GCD) extends Novel Class Discovery
(NCD)~\cite{ncd2,ncd3,ncd4} to unlabeled data containing both known
and novel categories~\cite{vaze2022generalized}. Early approaches
combine contrastive representation learning with semi-supervised
clustering, while SimGCD~\cite{wen2023parametric} introduces
parametric classification with self-distillation. Subsequent studies
improve category separation through information
maximization~\cite{chiaroni2023parametric}, prototypical
learning~\cite{zhang2024dual}, contrastive affinity learning with
auxiliary prompts~\cite{zheng2023promptcal}, spatial prompt
tuning~\cite{chen2024sptnet}, and contrastive mean-shift
learning~\cite{cms}. Recent advances further explore fine-grained
discrimination~\cite{selex,dai2025adaptive}, hyperbolic representation
learning~\cite{liu2025hyperbolic}, more effective utilization of
unlabeled data~\cite{cao2025allgcd}, and robustness to domain
shifts~\cite{wang2025hilo}. ProtoGCD~\cite{ma2025protogcd} jointly
models known and novel categories with unified prototypes and
adaptive pseudo-labeling to mitigate confirmation bias.
TAR~\cite{yang2026tar} addresses attention artifacts through
token-aware refinement, improving the use of discriminative local
cues for fine-grained discovery. These methods primarily improve
the organization and discriminability of visual representations.

Another line of research incorporates linguistic knowledge and
multimodal information into GCD. CLIP-GCD~\cite{may2023clip} and
TextGCD~\cite{zheng2024textual} exploit retrieved textual information,
with TextGCD additionally introducing cross-modality co-teaching.
MM-GCD~\cite{su2025multimodal} studies paired image--text
observations, while GET~\cite{wang2025get} synthesizes pseudo text
embeddings from visual features. CPT~\cite{add2} addresses ambiguity
in category definitions through consistent prompt tuning, and
GraphVL~\cite{solanki2024graphvl} preserves semantic neighborhood
structure through graph-enhanced vision--language modeling.
SSR$^2$-GCD~\cite{ssr2gcd} further investigates structured multimodal
representations through semi-supervised rate reduction.
ATCG~\cite{han2026learning} generates textual concepts for unlabeled
samples by analogy with labeled knowledge and combines them with
visual features to improve category separation.
Beyond vision--language learning, Cheng et al.~\cite{cheng2026multimodal}
study multimodal generalized defect category discovery using images
and point clouds, with defect-aware representations and calibrated
clustering.

These studies establish the value of complementary information for
category discovery. Our work focuses on the relative reliability of
appearance and geometry in paired 2D--3D object observations. Under
partial category supervision, uncertain predictions from either
modality can affect both the fused representation and the knowledge
transferred between branches. RaGCD therefore jointly regulates
modality contributions and cross-modal alignment according to
sample-wise reliability, with the aim of preserving known-category
discrimination while organizing novel categories.

\subsection{Multimodal Learning}

Multimodal learning combines heterogeneous information through
representation alignment, feature fusion, and cross-modal knowledge
transfer~\cite{baltrusaitis2018multimodal,liang2024foundations}.
CLIP~\cite{radford2021learning} demonstrates the transferability of
representations learned from image--text alignment. Extending this
paradigm to 3D understanding, ULIP~\cite{xue2023ulip} aligns point
cloud representations with image and language embeddings, while
ULIP-2~\cite{xue2024ulip2}, OpenShape~\cite{liu2023openshape}, and
Uni3D~\cite{zhou2024uni3d} investigate scalable multimodal
pre-training for transferable 3D representations.
Point-Bind~\cite{guo2024point} further connects point clouds with
multiple modalities. Complementary work studies image-to-point
representation transfer~\cite{zhang2023learning}, joint learning
from multi-view images and point clouds~\cite{add6}, and
task-specific image--point interactions for 3D
detection~\cite{vora2020pointpainting,add9}. These approaches support
applications ranging from supervised perception to cross-modal
retrieval and zero-shot recognition, but do not directly address
the GCD objective of discovering novel categories from partially
labeled data.

Multimodal learning also faces unequal learning progress and
variable input quality. OGM~\cite{peng2022balanced} and
PMR~\cite{fan2023pmr} alleviate modality imbalance through gradient
modulation and prototype-based regularization, respectively.
InfoReg~\cite{huang2025adaptive} regulates early-stage information
acquisition to reduce suppression by dominant modalities.
ARL~\cite{wei2025improving} instead motivates asymmetric modality
optimization through bias--variance analysis, highlighting that
equal modality dependence is not necessarily optimal.
For incomplete observations, BALM~\cite{nguyen2026balm} combines
feature calibration and gradient rebalancing to address unequal
modality missing rates. Complementing these optimization-oriented
approaches, TMC~\cite{han2021trusted} and
QMF~\cite{zhang2023provable} investigate uncertainty-aware evidence
integration and quality-aware fusion.
CoRiM~\cite{zou2026corim} further formulates dynamic fusion as
sample-wise conflict-risk minimization, jointly considering fused
uncertainty and inter-modal consistency. These studies establish
the importance of adapting modality treatment to learning
dynamics, input availability, and cross-modal conflict.

RaGCD investigates this issue in paired 2D--3D GCD, where labels are
unavailable for novel categories. In this setting, reliability
affects both modality allocation and the cross-modal interactions
used to learn category structure. Our reliability-aware modality
allocation adjusts channel-wise appearance and geometry
contributions, while reliability-conditioned alignment regulates
feature-, prediction-, and prototype-level knowledge transfer.
Together, they adapt multimodal integration to the supervision
constraints of category discovery.
\section{Method}
\label{sec:method}

\subsection{Problem Formulation}
\label{sec:problem}
We consider generalized category discovery (GCD) from paired images
and point clouds. The training data comprise a labeled set and an
unlabeled set:
\begin{equation}
\begin{aligned}
\mathcal D_L&=\{(x_i^I,x_i^P,y_i)\}_{i=1}^{N_L},\\
\mathcal D_U&=\{(x_i^I,x_i^P)\}_{i=1}^{N_U},
\end{aligned}
\end{equation}
where $x_i^I$ denotes one or more images of an object, $x_i^P$ its
point cloud, and $N_L$ and $N_U$ the corresponding set sizes.
Labels $y_i$ are available only for known categories. The unlabeled
set contains both known and novel categories. We denote the total
and known category counts by $C$ and $C_K$, with known categories
indexed first, and assume $C$ is available. The objective is to learn
a representation for recognizing known categories and discovering
novel categories without accessing unlabeled ground-truth labels.

\subsection{Framework Overview}
\label{sec:overview}
\begin{figure*}[t]
    \centering
    \includegraphics[width=\textwidth]{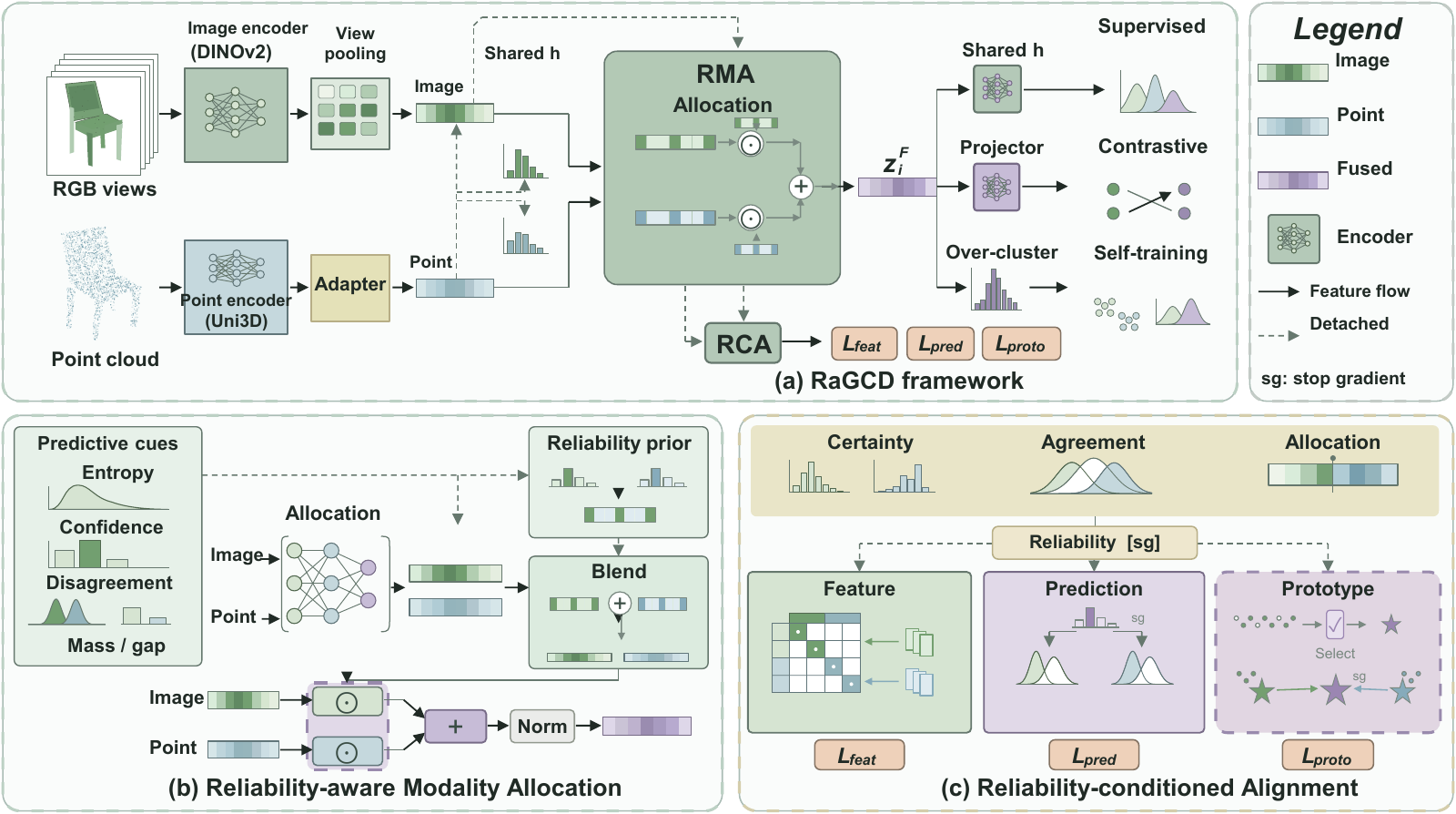}
    \caption{
    Overview of RaGCD for paired 2D--3D generalized category
    discovery. Modality-specific encoders extract appearance
    and geometric representations, with multiple image views
    aggregated before fusion. Reliability-aware modality
    allocation (RMA) adaptively determines channel-wise
    contributions. Reliability-conditioned alignment (RCA)
    supports cross-modal transfer at the feature, prediction,
    and prototype levels.
    }
    \label{fig:framework}
\end{figure*}

As illustrated in Fig.~\ref{fig:framework}, RaGCD combines two
modality-specific branches with reliability-aware modality allocation
(RMA) and reliability-conditioned alignment (RCA). Image views are
aggregated by attention pooling, and a learnable adapter maps
point-cloud features to the image feature dimension. The resulting
features $z_i^I,z_i^P\in\mathbb R^d$ are $\ell_2$-normalized.
A shared classification head produces distributions $p_i^I,p_i^P$
over the same $C$ output categories. RMA uses these predictions and
the features to construct $z_i^F$, from which the same head produces
$p_i^F$. Thus unimodal predictions are available before fusion;
computing the reliability cues does not require a circular prediction
step.

RMA regulates channel-wise modality contributions. RCA reuses
prediction statistics and allocation information to weight
cross-modal learning and select observations for prototype transfer.
The fused branch is trained with supervised and unsupervised discovery
objectives, including an auxiliary over-clustering head.


\subsection{Reliability-Aware Modality Allocation}
\label{sec:gate}
RMA adapts the contributions of appearance and geometry across
samples and channels. For each branch, normalized entropy and maximum
class probability are
\begin{equation}
e_i^m=-\frac{\sum_{k=1}^{C}p_{i,k}^m\log p_{i,k}^m}{\log C},
\qquad c_i^m=\max_k p_{i,k}^m,
\label{eq:prediction_statistics}
\end{equation}
where $m\in\{I,P,F\}$. For the two unimodal branches, we further use
known-category probability mass $b_i^m=\sum_{k=1}^{C_K}p_{i,k}^m$
and prediction disagreement
\begin{equation}
J_i=\tfrac12\sum_{m\in\{I,P\}}
D_{\mathrm{KL}}\!\left(p_i^m\middle\|\frac{p_i^I+p_i^P}{2}\right),
\end{equation}
where $D_{\mathrm{KL}}$ denotes Kullback--Leibler divergence.
Natural logarithms give $e_i^m\in[0,1]$ and $J_i\in[0,\log 2]$.

A lightweight network takes the concatenated features and eight
predictive cues, $(e_i^I,e_i^P,c_i^I,c_i^P,J_i,b_i^I,b_i^P,
|b_i^I-b_i^P|)$, and outputs positive channel scores
$s_i^I,s_i^P\in(0,1)^d$. It consists of two linear layers with
LayerNorm and ReLU between them and a final sigmoid; the hidden
width is $\max(d/2,128)$. Predictive cues are held fixed when
backpropagating through this network, while its feature inputs remain
trainable.

An entropy--confidence estimate provides a complementary prior:
\begin{equation}
\rho_i^m=\tfrac12(1-e_i^m+c_i^m),\qquad
\pi_i^m=\frac{\exp(2\rho_i^m)}{\sum_{n\in\{I,P\}}\exp(2\rho_i^n)}.
\end{equation}
For $m\in\{I,P\}$, the allocation scores and weights are
\begin{equation}
a_i^m=(1-\eta)s_i^m+\eta\pi_i^m,\qquad
w_i^m=\frac{a_i^m}{a_i^I+a_i^P},
\end{equation}
where $\eta\in[0,1]$ controls the prior contribution and each scalar
prior is broadcast across channels. The fused feature is
\begin{equation}
z_i^F=\operatorname{Norm}\!\left(
\sum_{m\in\{I,P\}}w_i^m\odot z_i^m\right),
\label{eq:reliability_fusion}
\end{equation}
where division and $\odot$ are element-wise and $\operatorname{Norm}$
is $\ell_2$ normalization. Numerical stabilization is omitted here.
These prediction-derived statistics guide allocation; they are
heuristic reliability cues, not calibrated probabilities of correctness.

\subsection{Reliability-Conditioned Alignment}
\label{sec:alignment}
Adaptive fusion does not guarantee a correct fused prediction.
RCA therefore controls each sample's relative contribution to
cross-modal learning. For an unlabeled sample, the reliability is
\begin{equation}
\begin{aligned}
\hat r_i={}&(1-e_i^F)\left(1-\frac{J_i}{\log 2}\right)
\frac{\rho_i^I+\rho_i^P}{2}\\
&\cdot\left(1-2\operatorname{Var}(w_i^P)\right)
\left(1-\left|\operatorname{Mean}(w_i^P)-\tfrac12\right|\right),\\
r_i={}&\max(\hat r_i,\alpha_{\min}).
\end{aligned}
\label{eq:alignment_weight}
\end{equation}
Mean and population variance are computed across feature channels.
The first three factors combine fused certainty, cross-modal agreement,
and average unimodal reliability. The last two factors reduce the
priority of dispersed channel allocations and strong single-modality
dependence for alignment. Each of these two factors lies in $[1/2,1]$;
thus $\hat r_i\in[0,1]$. The lower bound
$\alpha_{\min}\in[0,1]$ prevents uncertain samples from receiving
arbitrarily small weights. When RMA is disabled, its two allocation
factors are set to one.

Labeled samples receive a uniform positive weight. Reliability
weights are treated as constants during loss optimization. Feature
and prediction losses are evaluated separately on labeled and
unlabeled mini-batches and then summed. Their weight normalization
controls relative sample contributions; it does not guarantee a
smaller loss when all weights in a batch decrease together.

\paragraph{Feature-level alignment.}
Separate projection heads map $z_i^I$ and $z_i^P$ into a shared
contrastive space. Symmetric cross-modal InfoNCE uses paired objects
as positives and other objects in the batch as negatives:
\begin{equation}
\mathcal L_{\mathrm{feat}}=
\frac{\sum_i r_i\mathcal L_{\mathrm{con}}(z_i^I,z_i^P)}
{\max(\sum_i r_i,\epsilon)},
\end{equation}
where $\mathcal L_{\mathrm{con}}$ includes the two directions with
equal weight and operates on normalized projected features;
$\epsilon>0$ ensures numerical stability. Negatives are gathered
across devices in distributed training.

\paragraph{Prediction-level alignment.}
The fused prediction serves as a fixed soft target for both
unimodal branches:
\begin{equation}
\mathcal L_{\mathrm{pred}}=
\frac{\sum_i r_i\left[
D_{\mathrm{KL}}(p_i^F\|p_i^I)+D_{\mathrm{KL}}(p_i^F\|p_i^P)
\right]}{\max(\sum_i r_i,\epsilon)}.
\end{equation}
Reliability weighting reduces the relative influence of uncertain or
inconsistent samples on prediction transfer.

\paragraph{Prototype-level alignment.}
All branches use the fused pseudo-label $\hat y_i=\arg\max_k p_{i,k}^F$.
An unlabeled sample is retained only if $c_i^F\geq\tau_p$ and
$r_i\geq\tau_r$, where $\tau_p$ and $\tau_r$ are confidence and
reliability thresholds. For each retained pseudo-category, we compute
its prototype in each branch as the normalized $r_i$-weighted mean
of its selected features. Prototype construction pools samples across
devices. The unimodal prototypes are aligned to fixed fused targets:
\begin{equation}
\mathcal L_{\mathrm{proto}}=
\frac{1}{|\mathcal K|}\sum_{k\in\mathcal K}
\sum_{m\in\{I,P\}}\left[1-\cos(\mu_k^m,\mu_k^F)\right],
\end{equation}
where $\mathcal K$ is the set of valid pseudo-categories and
$\mu_k^m$ is the prototype of branch $m\in\{I,P,F\}$.
The loss is zero if no prototype is valid; the implementation also
skips construction when the total retained weight is below one.
Under the reported thresholds, labeled weights fall below $\tau_r$,
so labeled samples do not contribute prototypes.

The alignment objective is
\begin{equation}
\mathcal L_{\mathrm{align}}=
\mathcal L_{\mathrm{feat}}+\mathcal L_{\mathrm{pred}}
+\lambda_{\mathrm{proto}}\mathcal L_{\mathrm{proto}},
\end{equation}
where $\lambda_{\mathrm{proto}}$ controls prototype transfer. RMA
and RCA serve different purposes: a strong preference for one modality
can aid fusion while reducing that sample's priority for mutual
alignment. Agreement and confidence can nevertheless be high when
both modalities are wrong; RCA does not eliminate this failure mode.

\subsection{GCD Learning Objective}
\label{sec:gcd_objective}

For labeled samples, cross-entropy and supervised
contrastive learning preserve known-category discrimination:
\begin{equation}
\mathcal L_{\mathrm{lab}}=
-\frac{1}{B_L}\sum_{i=1}^{B_L}\log p_{i,y_i}^{K}
+\tfrac12\mathcal L_{\mathrm{supcon}},
\end{equation}
where $B_L$ is the labeled batch size and $p_i^K$ is the
fused prediction renormalized over the $C_K$ known
categories. $\mathcal L_{\mathrm{supcon}}$ applies supervised
contrastive learning to normalized projected fused
features, using same-class samples as positives and
excluding self-pairs.

For unlabeled samples, weakly and strongly augmented
images share the same point cloud. The discovery objective
combines instance contrastive learning, pseudo-label
supervision, and entropy regularization:
\begin{equation}
\begin{aligned}
\mathcal L_{\mathrm{unlab}}={}&
-\frac{1}{B_U}\sum_{i=1}^{B_U}
\log\frac{\exp((u_i^w)^\top u_i^s/\tau)}
{\sum_{j\in\mathcal S}
\exp((u_i^w)^\top u_j^s/\tau)}\\
&+\sum_{h\in\mathcal H}
\left[\mathcal L_{\mathrm{PL}}^h
-\tfrac12 H(\bar p^{h,w})\right],
\end{aligned}
\end{equation}
where $B_U$ is the unlabeled batch size, $u_i^w,u_i^s$
are normalized projected fused features, and $\tau$ is
the contrastive temperature. $\mathcal S$ contains the
strong-view candidates, gathered across devices during
distributed training. The active head set
$\mathcal H\subseteq\{\mathrm{std},\mathrm{over}\}$ includes
the standard head and, when enabled, the auxiliary
over-clustering head.

For each head $h$, weak-view predictions $p_i^{h,w}$
generate confidence-filtered hard pseudo-labels:
\begin{equation}
\mathcal L_{\mathrm{PL}}^h=
-\frac{1}{B_U}\sum_{i=1}^{B_U}
\mathbb I\left[\max_k p_{i,k}^{h,w}\geq\theta\right]
\log p_{i,\hat y_i^h}^{h,s},
\end{equation}
where $\hat y_i^h=\arg\max_k p_{i,k}^{h,w}$,
$p_i^{h,s}$ is the strong-view prediction, and
$\theta=0.7$. The mean weak-view prediction
$\bar p^{h,w}=B_U^{-1}\sum_i p_i^{h,w}$ is regularized
by $H(p)=-\sum_k p_k\log p_k$ to discourage category
collapse. The standard head has $C$ outputs, while
the over-clustering head has $\gamma C$ outputs,
with $\gamma=5$ for ModelNet40 and $\gamma=3$
for Washington RGB-D and ShapeNet-ViPC.

The complete objective during joint training is
\begin{equation}
\mathcal L=\mathcal L_{\mathrm{lab}}
+\mathcal L_{\mathrm{unlab}}
+\lambda_{\mathrm{align}}\mathcal L_{\mathrm{align}},
\end{equation}
where $\lambda_{\mathrm{align}}$ controls the overall
RCA contribution. RMA predictive cues and their prior
are detached from unimodal logits, while feature inputs
remain trainable. RCA weights, pseudo-label targets,
and fused alignment targets are treated as fixed during
their respective loss computations, whereas the entropy
regularizer remains differentiable.

\section{Experiments}
\label{sec:experiments}

\subsection{Experimental Setup}
\label{sec:exp_setup}

\paragraph{Datasets and GCD protocol.}
We evaluate RaGCD on ModelNet40~\cite{wu20153dshapenets},
Washington RGB-D~\cite{lai2011large}, and
ShapeNet-ViPC~\cite{zhang2021view}, covering synthetic CAD objects, real RGB-D observations, and partial geometry, respectively.
The known/novel category splits are 20/20, 26/25, and 7/6.
Unless otherwise specified, 50\% of the training samples in each
known category are labeled; the remaining known samples and all
novel samples form the unlabeled set. The default category partition
and labeled subset use seed 42 and are shared by the compared methods.
Table~\ref{tab:dataset_statistics} summarizes the training data and inputs.

ModelNet40 uses the official object-level train/test split, with four
rendered RGB views paired with each object's point cloud.
Washington RGB-D uses an instance-disjoint 80/20 split within each
category: all frames of a physical object remain in one partition,
and RGB crops are paired with point clouds by frame identifier.
The instance assignments are stored in a fixed JSON split file.
ShapeNet-ViPC uses an object-disjoint 80/20 split, grouping rendered
views and partial point clouds by the underlying ShapeNet object.
Partial point clouds are used as its default geometric input.
These grouping rules prevent observations of the same object from
appearing in both training and test partitions.

We sample 1,024 points for ModelNet40 and Washington RGB-D and
2,048 points for ShapeNet-ViPC. Each cloud is centered at its centroid
and divided by its maximum distance from the centroid.
Table~\ref{tab:dataset_statistics} also clarifies that the unlabeled
set spans the complete category set rather than novel categories alone.
The three datasets test different observation conditions: multiple rendered
views in ModelNet40, real object frames in Washington RGB-D, and partial
geometry in ShapeNet-ViPC. The point count and view count are therefore
specified per dataset rather than assumed to be identical across benchmarks.
Object- and instance-disjoint partitions are particularly relevant for
multi-view data, since visually similar observations of one object should
not appear on both sides of the train/test split.

\begin{table*}[t]
\centering
\small
\setlength{\tabcolsep}{5pt}
\caption{Training-set statistics and input configurations.}
\label{tab:dataset_statistics}
\begin{tabular}{lcccccll}
\toprule
\multirow{2}{*}{\textbf{Name}}
& \multicolumn{2}{c}{\textbf{Labeled Set}}
& \multicolumn{2}{c}{\textbf{Unlabeled Set}}
& \multirow{2}{*}{\textbf{Input}}
& \multirow{2}{*}{\textbf{Train/Test Split}}
& \multirow{2}{*}{\textbf{Domain}} \\
\cmidrule(lr){2-3}
\cmidrule(lr){4-5}
& \#Samples & \#Classes
& \#Samples & \#Classes
& & & \\
\midrule
ModelNet40
& 2,551 & 20
& 7,292 & 40
& 4 views + 1,024 pts
& official object-level
& Synthetic CAD \\

Washington RGB-D
& 610 & 26
& 1,770 & 51
& 1 view + 1,024 pts
& instance-level 80/20
& Real RGB-D \\

ShapeNet-ViPC
& 7,932 & 7
& 22,731 & 13
& 1 view + 2,048 pts
& object-level 80/20
& Synthetic partial 3D \\
\bottomrule
\end{tabular}
\end{table*}

\paragraph{Evaluation metrics.}
We extract normalized fused features from the test partition and apply
$k$-means with the number of clusters set to the total category count,
except in the cluster-number sensitivity experiment. The implementation
uses \texttt{random\_state=0} and \texttt{n\_init=10}.
Hungarian matching~\cite{kuhn1955hungarian} aligns cluster assignments
with ground-truth categories. Clustering accuracy is
\begin{equation}
\mathrm{ACC}
=\max_{\pi}\frac{1}{N}\sum_{i=1}^{N}
\mathbb{I}\left[y_i=\pi(\hat y_i)\right],
\end{equation}
where $N$ is the number of evaluated samples, $\hat y_i$ is a cluster
assignment, and $\pi$ is a one-to-one matching between cluster and
category indices. We report accuracy over all (All), known (Old),
and novel (New) categories. All accuracy and H-score values are reported in percentages.
For alignment ablations, we additionally report their harmonic mean:
\begin{equation}
H=\frac{2\,\mathrm{Old}\,\mathrm{New}}
{\mathrm{Old}+\mathrm{New}}.
\end{equation}
For each representation, a single matching is computed over all test
samples; Old and New accuracies reuse that matching on their respective
subsets.

\paragraph{Implementation details.}
We use pretrained DINOv2 ViT-B/14~\cite{oquab2023dinov2} and
Uni3D-B~\cite{zhou2024uni3d} as the image and point-cloud encoders.
A learnable adapter maps point-cloud features to the 768-dimensional
image feature space, and attention pooling aggregates image views.
Most backbone parameters are frozen. The final DINOv2 transformer
block is fine-tuned, and the final Uni3D block is additionally unfrozen
in the multimodal alignment stage. Task-specific trainable components
include view aggregation, the modality adapter, RMA, classification
heads, and contrastive and alignment projectors.

The reported training configuration uses AdamW for 200 epochs, with
learning rates of $10^{-3}$ for newly initialized modules and $10^{-5}$
for unfrozen backbone parameters. Each labeled and unlabeled loader
uses a batch size of 32 per GPU and eight workers. Automatic mixed
precision is enabled. Weak image augmentation uses random resized
cropping and horizontal flipping; strong augmentation additionally
uses color jitter, grayscale conversion, and Gaussian blur.
Images are processed at $224\times224$ with ImageNet normalization.
Weak and strong image observations share the same paired point cloud.

The contrastive temperature is 0.07. The RMA reliability-prior strength
is $\eta=0.25$. For RCA, we use $\alpha_{\min}=0.1$,
$\lambda_{\mathrm{align}}=0.75$, and feature, prediction, and prototype
coefficients of 1.0, 1.0, and 0.5, respectively. Labeled samples receive
a uniform alignment weight of 0.3; this constant cancels in each
separately normalized instance loss and is below the prototype
reliability threshold. Prototype construction selects
unlabeled samples using confidence and reliability thresholds of
$\tau_p=0.8$ and $\tau_r=0.35$, respectively.

The reported training configuration also uses a pseudo-label confidence
threshold of 0.7 and an over-clustering factor of 5 for ModelNet40 and
3 for Washington RGB-D and ShapeNet-ViPC.

All methods share the category partitions and evaluation protocol.
The comparison includes methods with different input modalities and
pretraining sources; controlled component ablations use the same
DINOv2 and Uni3D backbones.

\subsection{Comparison with Representative Methods}
\label{sec:sota_comparison}

We compare RaGCD with GCD~\cite{vaze2022generalized},
HiLo~\cite{wang2025hilo}, CPF-GCD~\cite{cpfgcd},
SSR$^2$-GCD~\cite{ssr2gcd}, MM-GCD~\cite{su2025multimodal},
GET~\cite{wang2025get}, MM-Point~\cite{mmpoint},
SCA-PVNet~\cite{scapvnet}, and ClustView~\cite{clustview}.
MM-Point, SCA-PVNet, and ClustView are adapted to the GCD setting and
evaluated through their learned representations using the same
clustering protocol.

Table~\ref{tab:main_comparison} reports All, Old, and New accuracy;
bold and underlined entries identify the best and second-best result
in each column, respectively. RaGCD achieves the highest All and New
accuracy on all three benchmarks, while its Old accuracy is highest on
ModelNet40 and Washington RGB-D.

On ModelNet40, RaGCD obtains 84.20\% All and 73.70\% New accuracy.
The respective second-best methods are MM-Point for All and CPF-GCD
for New, giving gains of 3.62 and 6.01 percentage points. Its Old
accuracy of 95.26\% is close to MM-Point's 95.16\%, so the larger
difference occurs in novel-category discovery. On Washington RGB-D,
RaGCD achieves 88.06\% All, 92.26\% Old, and 83.87\% New accuracy.
Relative to CPF-GCD, the second-best method on all three metrics,
the gains are 1.26, 0.14, and 2.38 points. These comparisons indicate
that the observed improvement is more pronounced for novel categories
than for already well-recognized known categories on these two datasets.

On ShapeNet-ViPC, RaGCD reaches 83.73\% All and 70.13\% New accuracy,
improving the respective second-best results by 4.46 and 0.87 points.
MM-Point has a higher Old accuracy of 97.08\%, compared with RaGCD's
96.40\%, but a lower New accuracy of 58.75\%. This comparison illustrates
why Old accuracy alone does not adequately characterize GCD performance.
Taken together, the results support RaGCD across rendered images, real
RGB-D observations, and partial geometric inputs. The following controlled
ablations examine the contributions of its individual components.

\begin{table*}[t]
\centering
\small
\setlength{\tabcolsep}{4pt}
\renewcommand{\arraystretch}{1.05}

\caption{Comparison with representative methods on three benchmarks.}
\label{tab:main_comparison}

\begin{tabularx}{\textwidth}{l*{9}{>{\centering\arraybackslash}X}}
\toprule
\multirow{2}{*}{\textbf{Method}}
& \multicolumn{3}{c}{\textbf{ModelNet40}}
& \multicolumn{3}{c}{\textbf{Washington RGB-D}}
& \multicolumn{3}{c}{\textbf{ShapeNet-ViPC}} \\
\cmidrule(lr){2-4}
\cmidrule(lr){5-7}
\cmidrule(lr){8-10}
& \textbf{All} & \textbf{Old} & \textbf{New}
& \textbf{All} & \textbf{Old} & \textbf{New}
& \textbf{All} & \textbf{Old} & \textbf{New} \\
\midrule

GCD~\cite{vaze2022generalized}
& 77.35 & 94.09 & 61.45
& 73.28 & 80.12 & 66.45
& 76.46 & \underline{96.62} & 54.85 \\

HiLo~\cite{wang2025hilo}
& 77.27 & 87.69 & 67.38
& 79.35 & 86.45 & 72.26
& 77.78 & 96.34 & 57.85 \\

CPF-GCD~\cite{cpfgcd}
& 78.81 & 90.52 & \underline{67.69}
& \underline{86.80} & \underline{92.12} & \underline{81.49}
& 79.00 & 88.08 & \underline{69.26} \\

MM-GCD~\cite{su2025multimodal}
& 74.19 & 89.60 & 59.56
& 81.94 & 86.45 & 77.42
& 73.10 & 93.67 & 51.04 \\

GET~\cite{wang2025get}
& 65.84 & 81.28 & 51.18
& 78.71 & 86.45 & 70.97
& 69.15 & 93.77 & 42.74 \\

SSR$^2$-GCD~\cite{ssr2gcd}
& 66.21 & 82.86 & 50.39
& 80.32 & 80.00 & 80.65
& \underline{79.27} & 89.67 & 68.11 \\

MM-Point~\cite{mmpoint}
& \underline{80.58} & \underline{95.16} & 66.73
& 85.81 & 90.97 & 80.65
& 78.58 & \textbf{97.08} & 58.75 \\

SCA-PVNet~\cite{scapvnet}
& 68.27 & 88.29 & 49.29
& 79.35 & 89.03 & 69.68
& 75.02 & 91.35 & 57.50 \\

ClustView~\cite{clustview}
& 66.37 & 90.43 & 43.52
& 80.97 & 91.61 & 70.32
& 78.72 & 95.27 & 60.96 \\

\midrule
\rowcolor{blue!6}
\textbf{RaGCD (Ours)}
& \textbf{84.20} & \textbf{95.26} & \textbf{73.70}
& \textbf{88.06} & \textbf{92.26} & \textbf{83.87}
& \textbf{83.73} & 96.40 & \textbf{70.13} \\

\bottomrule
\end{tabularx}
\end{table*}

\subsection{Ablation Studies}
\label{sec:ablation}

\paragraph{Effect of RMA and RCA.}

Table~\ref{tab:component_ablation} evaluates RMA and RCA on Washington
RGB-D. The fixed-fusion baseline assigns equal contributions to both
modalities and disables RCA. It obtains 78.61\% All, 91.65\% Old,
and 65.60\% New accuracy, leaving a substantial gap between known-
and novel-category performance.

Adding RMA increases All accuracy by 6.05 percentage points and New
accuracy by 14.42 points, although Old accuracy decreases from 91.65\%
to 89.31\%. Adaptive allocation therefore changes the recognition--discovery
trade-off rather than uniformly improving every metric. Adding RCA alone
to fixed fusion yields 83.64\% All and 75.57\% New accuracy, showing that
cross-modal transfer also provides gains without learned allocation.

The full model achieves the highest accuracy on all three metrics.
Relative to RMA alone, RCA adds 3.40 points on All, 2.95 on Old, and
3.85 on New, recovering the known-category decrease while further
improving discovery. Relative to the fixed-fusion baseline, the All and
New gains are 9.45 and 18.27 points. The combined improvement supports
using allocation and alignment together rather than relying on either
component alone.

When RMA is disabled, the allocation-dependent factors in RCA are
set to one, as defined in Section~\ref{sec:alignment}.

\begin{table}[t]
\centering
\small
\setlength{\tabcolsep}{5.0pt}
\renewcommand{\arraystretch}{1.08}
\caption{Ablation of RMA and RCA on Washington RGB-D.}
\label{tab:component_ablation}
\begin{tabular}{lccccc}
\toprule
\textbf{Setting}
& \textbf{RMA}
& \textbf{RCA}
& \textbf{All}
& \textbf{Old}
& \textbf{New} \\
\midrule
Fixed-fusion baseline
& $\times$
& $\times$
& 78.61
& 91.65
& 65.60 \\

RMA only
& $\checkmark$
& $\times$
& 84.66
& 89.31
& 80.02 \\

RCA only
& $\times$
& $\checkmark$
& 83.64
& 91.72
& 75.57 \\

\rowcolor{blue!6}
\textbf{RaGCD (Full)}
& $\checkmark$
& $\checkmark$
& \textbf{88.06}
& \textbf{92.26}
& \textbf{83.87} \\
\bottomrule
\end{tabular}
\end{table}

\paragraph{Comparison of fusion strategies.}

We compare fusion strategies using the same DINOv2 and Uni3D
representations, discovery heads, and training objectives on Washington
RGB-D. RCA is disabled in this experiment so that the comparison isolates
fusion. Table~\ref{tab:fusion_strategy} includes fixed averaging,
concatenation, attention-based fusion, scalar weighting, and RMA.

Concatenation obtains the highest conventional All accuracy of 83.23\%
and the highest Old accuracy of 92.90\%. Scalar weighting achieves the
highest conventional New accuracy of 74.19\%. RMA reaches 84.66\% All
and 80.02\% New, improving these respective alternatives by 1.43 and
5.83 percentage points. Compared with concatenation, RMA has a lower
Old accuracy but a 6.47-point higher New accuracy, indicating a different
balance between recognition and discovery.

The gain over fixed averaging is especially pronounced on New, at
14.42 points. Learned fusion alternatives also outperform fixed averaging,
but their different Old/New trade-offs show that adaptivity alone does
not yield identical behavior. RMA's strongest advantage in this comparison
is its novel-category performance; the next ablation examines how its
predictive reliability inputs contribute to that behavior.


\begin{table}[t]
\centering
\small
\setlength{\tabcolsep}{5.5pt}
\renewcommand{\arraystretch}{1.08}
\caption{Comparison of fusion strategies on Washington RGB-D.}
\label{tab:fusion_strategy}

\begin{tabular}{lccc}
\toprule
\textbf{Fusion Strategy}
& \textbf{All}
& \textbf{Old}
& \textbf{New} \\
\midrule

Fixed Mean
& 78.61
& \underline{91.65}
& 65.60 \\

Concatenation
& \underline{83.23}
& \textbf{92.90}
& 73.55 \\

Attention Gate
& 80.00
& 87.74
& 72.26 \\

Scalar Gate
& 82.90
& 91.61
& \underline{74.19} \\

\midrule

\rowcolor{blue!6}
\textbf{RMA}
& \textbf{84.66}
& 89.31
& \textbf{80.02} \\

\bottomrule
\end{tabular}
\end{table}

\paragraph{Effect of predictive reliability cues.}

Table~\ref{tab:signal_ablation} removes one predictive input group from
RMA at a time on Washington RGB-D. All variants retain the
entropy--confidence prior with $\eta=0.25$. The experiment therefore
tests each cue's direct contribution to the allocation network, rather
than removing every use of that statistic throughout the framework.

The full combination gives the highest All, Old, and New accuracy.
Removing confidence reduces All and New by 1.54 and 2.59 percentage
points, respectively, giving the largest New decrease among these
ablations. Removing JS disagreement reduces all three metrics by 1.94
points, producing the largest All decrease. Confidence describes each
branch's prediction strength, whereas disagreement describes the relation
between their predictions; the different degradation patterns are
consistent with these cues supplying complementary information.

Removing entropy reduces All and New by 0.70 and 1.29 points, while
removing known-category mass reduces them by 0.97 and 1.29 points.
The smaller entropy-input change should be interpreted with the retained
prior in mind, since entropy remains available through that path. The
results support the combined cue set, without implying that any one cue
is a calibrated measure of prediction correctness.


\begin{table}[t]
\centering
\caption{Ablation of predictive reliability cues on Washington RGB-D.}
\label{tab:signal_ablation}
\begin{tabular}{lccc}
\toprule
Setting & All & Old & New \\
\midrule
w/o Entropy Input
& 87.36 & 92.16 & 82.58 \\
w/o Confidence Input
& 86.52 & 91.76 & 81.28 \\
w/o JS Input
& 86.12 & 90.32 & 81.93 \\
w/o Known-class Mass Input
& 87.09 & 91.62 & 82.58 \\
\rowcolor{blue!6}
Full signals
& \textbf{88.06} & \textbf{92.26} & \textbf{83.87} \\
\bottomrule
\end{tabular}
\end{table}

\paragraph{Effect of reliability conditioning.}

Table~\ref{tab:reliability_conditioning} compares RMA without RCA,
the reported uniform-weight RCA variant ($r_i=1$), and full RCA
on Washington RGB-D. Uniform RCA obtains 84.81\% All,
90.91\% Old, and 78.62\% New accuracy. Relative to RMA alone,
it improves All by 0.15 points and Old by 1.60 points, while New
decreases by 1.40 points. Adding alignment therefore does not
uniformly improve known- and novel-category performance in this comparison.

Full RCA exceeds the uniform variant by 3.25, 1.35, and 5.25
percentage points on All, Old, and New, respectively. These results
support the complete reliability-conditioned configuration. They do not
separately establish the contributions of instance weighting and prototype
selection, nor establish that the reliability score ranks pseudo-label
correctness. The $\alpha_{\min}=1$ sensitivity setting also changes
prototype eligibility and is not used here as an interchangeable control.

\begin{table}[t]
\centering
\caption{Reliability conditioning on Washington RGB-D.}
\label{tab:reliability_conditioning}
\begin{tabular}{lccc}
\toprule
Alignment setting & All & Old & New \\
\midrule
RMA only, w/o RCA
& 84.66 & 89.31 & 80.02 \\
+ Uniform RCA, $r_i=1$
& 84.81 & 90.91 & 78.62 \\
\rowcolor{blue!6}
+ Reliability-conditioned RCA
& \textbf{88.06} & \textbf{92.26} & \textbf{83.87} \\
\bottomrule
\end{tabular}
\end{table}

\paragraph{Effect of alignment objectives.}

We retain the same RMA module and selectively enable feature, prediction,
and prototype alignment on Washington RGB-D
(Table~\ref{tab:alignment_ablation}). H-score is the harmonic mean of
Old and New accuracy defined in the evaluation protocol, providing a
complement to All accuracy when the two groups respond differently.

Among individual objectives, prototype alignment achieves the highest
All, New, and H-score, improving them over RMA alone by 1.95, 2.47,
and 2.00 percentage points. Feature alignment improves Old accuracy from
89.31\% to 91.08\%, but New decreases slightly from 80.02\% to
79.79\%. Prediction alignment instead improves both groups, reaching
89.84\% Old and 81.36\% New. Thus, the three objectives do not provide
interchangeable supervision: instance correspondence, prediction transfer,
and pseudo-category structure affect the discovery representation differently.

The two-objective combinations improve on the individual settings in All
accuracy. Feature--prototype alignment achieves the highest New accuracy
within this group, while prediction--prototype alignment gives the highest
All, Old, and H-score. The full combination obtains 88.06\% All,
83.87\% New, and an H-score of 87.87\%. Its Old accuracy is slightly
below prediction--prototype alignment, but the additional New gain yields
the highest H-score. This pattern supports the complementary use of all
three objectives rather than selecting a term using Old accuracy alone.

\begin{table}[t]
\centering
\caption{Ablation of RCA objectives on Washington RGB-D.}
\label{tab:alignment_ablation}
\resizebox{\linewidth}{!}{
\begin{tabular}{lcccc}
\toprule
Alignment setting & All & Old & New & H-score \\
\midrule
RMA only, w/o RCA
& 84.66 & 89.31 & 80.02 & 84.41 \\
Feature only
& 85.43 & 91.08 & 79.79 & 85.06 \\
Prediction only
& 85.59 & 89.84 & 81.36 & 85.39 \\
Prototype only
& 86.61 & 90.73 & 82.49 & 86.41 \\
Feature + Prediction
& 86.66 & 91.37 & 81.95 & 86.40 \\
Feature + Prototype
& 86.96 & 90.79 & 83.12 & 86.79 \\
Prediction + Prototype
& 87.74 & \textbf{92.44} & 83.06 & 87.50 \\
\rowcolor{blue!6}
Feature + Prediction + Prototype
& \textbf{88.06} & 92.26 & \textbf{83.87}
& \textbf{87.87} \\
\bottomrule
\end{tabular}
}
\end{table}

\paragraph{Effect of backbone choices.}
\label{sec:3.2}

Table~\ref{tab:backbone_ablation} evaluates alternative image and
point-cloud encoders on ModelNet40 under the same GCD protocol and
includes ULIP-2 as a pretrained representation baseline. The comparison
examines how the quality of the underlying representations affects the
framework's recognition and discovery performance.

With Point-BERT fixed, replacing DINOv1 with DINOv2 improves All and
New accuracy by 2.83 and 3.95 percentage points. With Uni3D fixed,
the corresponding gains are 5.92 and 9.32 points. Replacing Point-BERT
with Uni3D also improves All and New for either image encoder. Under
DINOv1, however, Old decreases from 93.43\% to 92.93\%, so the
geometric encoder change does not uniformly improve every metric.

DINOv2 + Uni3D yields the highest All, Old, and New values. Relative
to ULIP-2, it improves All by 2.79 points and New by 5.32 points.
The results demonstrate that RaGCD accommodates multiple encoder pairs
and that pretraining remains influential. The contribution of RMA and
RCA themselves is established by the matched component ablations above,
rather than by differences between pretrained encoder configurations.


\begin{table}[t]
\centering
\small
\setlength{\tabcolsep}{3.5pt}
\caption{Comparison of encoder configurations on ModelNet40.}
\label{tab:backbone_ablation}
\begin{tabular}{lccc}
\toprule
\textbf{Backbone / Representation} & \textbf{All} & \textbf{Old} & \textbf{New} \\
\midrule
DINOv1 + Point-BERT & 77.64 & 93.43 & 62.64 \\
DINOv1 + Uni3D      & 78.28 & 92.93 & 64.38 \\
DINOv2 + Point-BERT & 80.47 & 95.09 & 66.59 \\
ULIP-2              & 81.41 & 95.14 & 68.38 \\
\rowcolor{blue!6}
DINOv2 + Uni3D      & \textbf{84.20} & \textbf{95.26} & \textbf{73.70} \\
\bottomrule
\end{tabular}
\end{table}

\subsection{Analysis of Cross-Modal Reliability}
\label{sec:reliability_analysis}

\paragraph{Cross-modal reliability mismatch.}

We analyze predictions from the fixed-fusion baseline on ShapeNet-ViPC
to determine whether appearance and geometry provide equally dependable
evidence. The baseline uses the same backbones and GCD training protocol
as RaGCD, with equal modality contributions and no RCA. We compute JS
divergence between the unimodal prediction distributions and classify
samples as image-only correct, point-only correct, both correct, or both
wrong. Ground-truth labels are used only for this post-hoc diagnostic.

\begin{figure}[t]
\centering
\includegraphics[width=\columnwidth]{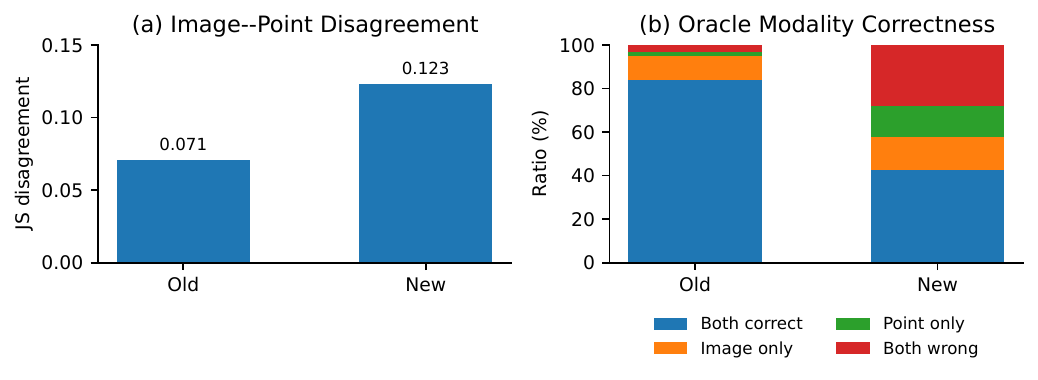}
\caption{Image--point disagreement and correctness on ShapeNet-ViPC using the fixed-fusion baseline. Ground-truth labels are used only for post-hoc analysis.}
\label{fig:mismatch_analysis}
\end{figure}

Fig.~\ref{fig:mismatch_analysis} reports mean JS divergence of 0.071
for known samples and 0.123 for novel samples. The larger disagreement
on novel categories indicates that the two branches organize these
observations less consistently. Disagreement alone does not identify
which branch is correct, so the correctness breakdown supplies a
complementary perspective. Both image-only-correct and point-only-correct
cases occur, showing that neither modality is uniformly preferable.

The higher both-wrong proportion on novel samples further shows that
selecting one modality cannot resolve every error. These findings motivate
both sample-dependent allocation and selective cross-modal transfer:
the former can favor useful evidence when branch behavior differs,
while the latter limits the influence of uncertain transfer targets.
Consistent but incorrect predictions remain a limitation: disagreement
cannot by itself detect errors shared by the two branches.


\paragraph{Response to modality degradation.}

We test whether RMA responds to changes in input quality using a fixed
checkpoint on Washington RGB-D. One modality is corrupted at a time
without retraining, and image, point-cloud, and fused representations
are evaluated under the same clustering protocol. This separates the
allocation response to degraded evidence from adaptation through
additional parameter updates.

\begin{figure}[t]
\centering
\includegraphics[width=\columnwidth]{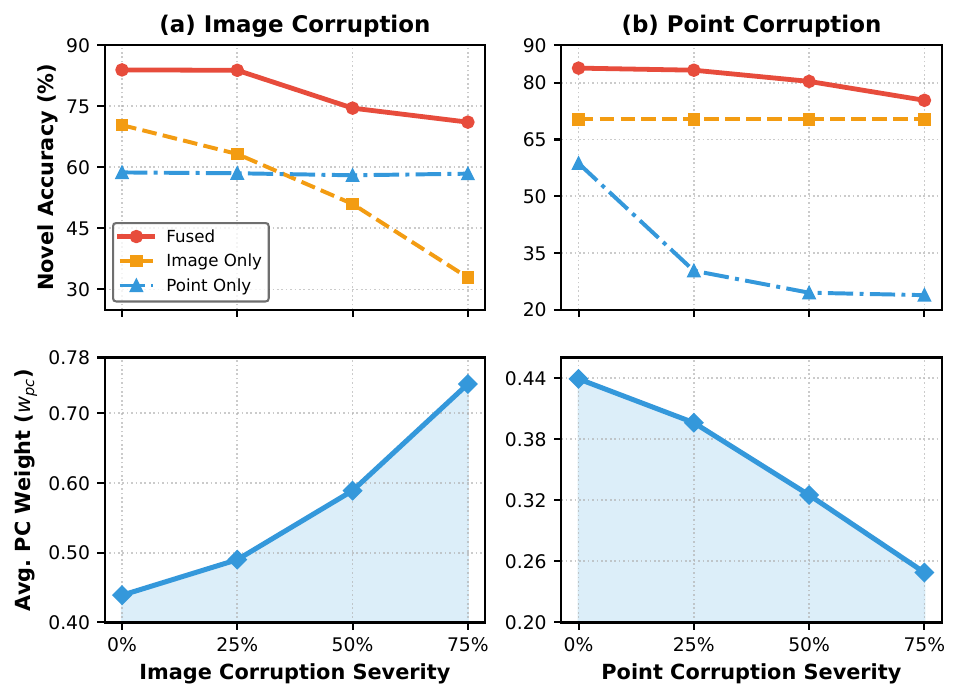}
\caption{Inference-time modality corruption on Washington RGB-D. One modality is corrupted at a time; curves show clustering accuracy and mean point-cloud allocation.}
\label{fig:modality_degradation}
\end{figure}

As shown in Fig.~\ref{fig:modality_degradation}, corruption primarily
reduces the performance of the affected unimodal branch. Under image
corruption, the average point-cloud allocation rises from 0.439 to 0.742,
a change of 0.303 toward geometric information. Under point-cloud
corruption, it falls to 0.249, a change of 0.190 toward appearance.
The opposing directions support a response to the affected modality
rather than a constant preference for one branch.

The fused representation remains more stable than the corrupted
unimodal representation in these tests. Allocation changes and accuracy
changes together provide more informative evidence than either measure
alone: changing weights shows a response, while retained accuracy shows
that the response accompanies useful predictions. These observations
support adaptation under the tested corruptions, without implying that
the correct modality can be identified for every individual sample.


\subsection{Robustness and Sensitivity}
\label{sec:sensitivity}

\paragraph{Random seeds and category partitions.}
\label{sec:seed_robustness}

We first vary the training seed while holding the known/novel category
partition and train/test split fixed. Table~\ref{tab:seed_stability}
reports seeds 1, 42, and 100, with seed 42 corresponding to the main
comparison. Mean $\pm$ standard deviation is calculated across these
three runs using the sample standard deviation.

Mean All accuracy is 84.01\% on ModelNet40, 88.03\% on Washington
RGB-D, and 83.70\% on ShapeNet-ViPC. The respective standard deviations
are 0.36, 0.39, and 0.37 percentage points. New accuracy is similarly
stable, with standard deviations of 0.33, 0.33, and 0.27 points.
The seed-42 results are close to their corresponding means and are not
the highest All result on any of the three datasets.

These runs provide evidence that the reported configuration is not
dependent on one favorable training initialization. Because the category
partition remains fixed, this experiment measures training variation;
the following comparison separately examines changes in category
composition. Three runs summarize the observed variability but do not
by themselves establish statistical significance against every baseline.

\begin{table*}[t]
\centering
\small
\setlength{\tabcolsep}{3.5pt}
\renewcommand{\arraystretch}{1.05}
\caption{Performance across training seeds on three benchmarks.}
\label{tab:seed_stability}

\resizebox{\textwidth}{!}{%
\begin{tabular}{lccccccccc}
\toprule
\multirow{2}{*}{\textbf{Seed}}
& \multicolumn{3}{c}{\textbf{ModelNet40}}
& \multicolumn{3}{c}{\textbf{Washington RGB-D}}
& \multicolumn{3}{c}{\textbf{ShapeNet-ViPC}} \\
\cmidrule(lr){2-4}
\cmidrule(lr){5-7}
\cmidrule(lr){8-10}
& \textbf{All} & \textbf{Old} & \textbf{New}
& \textbf{All} & \textbf{Old} & \textbf{New}
& \textbf{All} & \textbf{Old} & \textbf{New} \\
\midrule

1
& 83.60 & 94.72 & 73.05
& 88.41 & 92.71 & 84.12
& 83.31 & 95.84 & 69.86 \\

42
& 84.20 & 95.26 & 73.70
& 88.06 & 92.26 & 83.87
& 83.73 & 96.40 & 70.13 \\

100
& 84.23 & 95.61 & 73.43
& 87.63 & 91.81 & 83.46
& 84.05 & 96.78 & 70.39 \\

\midrule
\rowcolor{blue!6}
\textbf{Mean $\pm$ Std.}
& $84.01 \pm 0.36$
& $95.20 \pm 0.45$
& $73.39 \pm 0.33$
& $88.03 \pm 0.39$
& $92.26 \pm 0.45$
& $83.82 \pm 0.33$
& $83.70 \pm 0.37$
& $96.34 \pm 0.47$
& $70.13 \pm 0.27$ \\

\bottomrule
\end{tabular}%
}
\end{table*}

\label{sec:partition_robustness}

We separately vary the known/novel category partition on Washington
RGB-D using class seeds 42, 123, and 2026, retaining 26 known and
25 novel categories. Table~\ref{tab:partition_robustness} reports mean
All, Old, and New accuracy of 89.46\%, 93.73\%, and 84.90\%, with
sample standard deviations of 1.22, 1.28, and 1.07 percentage points.

All accuracy ranges from 88.06\% to 90.32\%, while New ranges from
83.87\% to 86.00\%. The default class partition has the lowest All
and New values among these three, so its use in the main comparison
does not select the strongest observed partition. The variation is
larger than in the training-seed experiment, consistent with category
composition presenting a different source of difficulty. RaGCD maintains
comparable performance across the tested compositions, although this
limited set does not cover all possible known/novel partitions.


\begin{table}[t]
\centering
\caption{Performance across category partitions on Washington RGB-D.}
\label{tab:partition_robustness}

\small
\setlength{\tabcolsep}{5pt}
\renewcommand{\arraystretch}{1.05}

\begin{tabular}{@{}c|ccc@{}}
\toprule
Class Seed & All & Old & New \\
\midrule
42   & 88.06 & 92.26 & 83.87 \\
123  & 90.32 & 94.38 & 86.00 \\
2026 & 90.00 & 94.55 & 84.83 \\
\midrule
\rowcolor{blue!6}
Mean $\pm$ Std.
& $89.46 \pm 1.22$
& $93.73 \pm 1.28$
& $84.90 \pm 1.07$ \\
\bottomrule
\end{tabular}
\end{table}

\paragraph{Labeled ratio.}
\label{sec:label_ratio}

\begin{figure}[t]
\centering
\includegraphics[width=0.95\columnwidth]{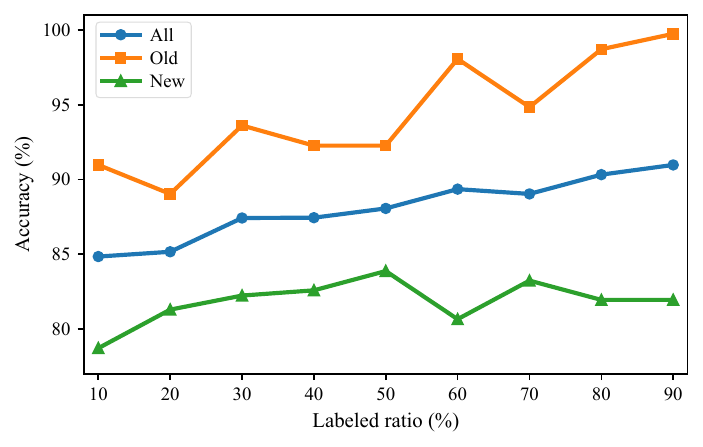}
\caption{Effect of the labeled ratio on Washington RGB-D.}
\label{fig:label_ratio}
\end{figure}

Fig.~\ref{fig:label_ratio} varies the proportion of labeled known-class
training samples on Washington RGB-D. All accuracy increases from
84.84\% at 10\% labels to 90.97\% at 90\% labels, and Old accuracy
reaches 99.74\% at the latter setting. More known-class annotations
therefore primarily strengthen recognition of the supervised categories
in this experiment.

New accuracy follows a different pattern. It rises from 78.71\% at
10\% labels to its observed maximum of 83.87\% at 50\%, but does not
continue increasing with additional known-class labels. An increase
in All accuracy at high labeled ratios should therefore not be
interpreted as a corresponding improvement in novel-category discovery.
The distinction is relevant to GCD, where additional labels describe
known categories rather than directly supervising the novel categories.

The default 50\% setting achieves the highest observed New accuracy
while using fewer annotations than the 90\% setting. The results support
reporting Old and New separately when assessing annotation efficiency,
since a single overall score would obscure their different responses.

\paragraph{Assumed category number.}

Table~\ref{tab:k_sensitivity} varies the assumed category count on
Washington RGB-D, whose ground-truth count is $K_{\mathrm{GT}}=51$.
The ratio column denotes the nominal multiples of this count, rounded
to the integer values of $K$ used in the experiment. The highest All,
Old, and New accuracies are obtained at the true category count.

At $K=38$, All accuracy falls from 88.06\% to 73.58\%, and New falls
from 83.87\% to 59.52\%. The larger New decrease is consistent with
insufficient cluster capacity forcing distinct novel groups to share
assignments. Moderate over-estimation has a smaller effect: at $K=64$,
All and New decrease by 2.65 and 2.03 percentage points, respectively.

Further over-estimation also degrades performance, with All and New
reaching 75.02\% and 66.96\% at $K=102$. Additional clusters can
fragment semantic categories, although the effect on accuracy also
depends on how unmatched clusters are scored. The tested results
indicate greater tolerance to moderate over-estimation than to
under-estimation; they do not remove the need to specify the category
count and matching protocol used for evaluation.


\begin{table}[t]
\centering
\small
\setlength{\tabcolsep}{5.5pt}
\renewcommand{\arraystretch}{1.08}
\caption{Sensitivity to the assumed category count $K$ on Washington RGB-D.}
\label{tab:k_sensitivity}
\begin{tabular}{ccccc}
\toprule
\textbf{Assumed $K$}
& \textbf{Ratio}
& \textbf{All}
& \textbf{Old}
& \textbf{New} \\
\midrule
38
& $0.75K_{\mathrm{GT}}$
& 73.58
& 87.68
& 59.52 \\

\rowcolor{blue!6}
51
& $1.00K_{\mathrm{GT}}$
& \textbf{88.06}
& \textbf{92.26}
& \textbf{83.87} \\

64
& $1.25K_{\mathrm{GT}}$
& \underline{85.41}
& \underline{88.98}
& \underline{81.84} \\

77
& $1.50K_{\mathrm{GT}}$
& 83.40
& 87.68
& 79.13 \\

89
& $1.75K_{\mathrm{GT}}$
& 78.66
& 87.68
& 69.66 \\

102
& $2.00K_{\mathrm{GT}}$
& 75.02
& 83.10
& 66.96 \\
\bottomrule
\end{tabular}
\end{table}

\paragraph{Minimum alignment weight.}

\begin{figure}[t]
    \centering
    \includegraphics[width=0.95\linewidth]
    {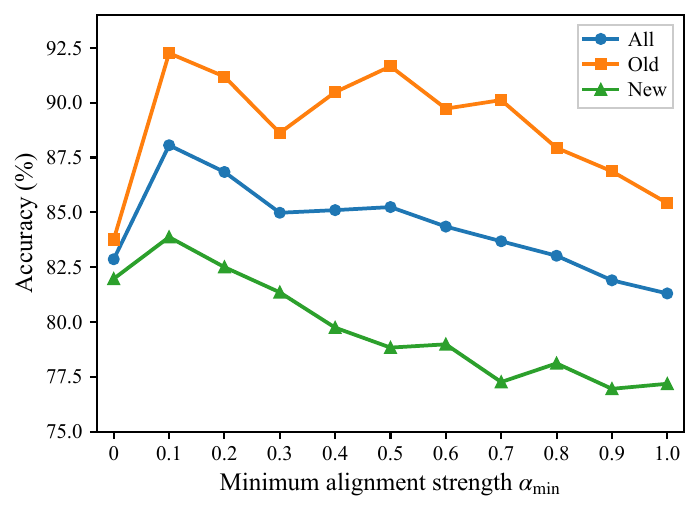}
    \caption{Effect of $\alpha_{\min}$ on Washington RGB-D.}
    \label{fig:alpha_min_sensitivity}
\end{figure}

Fig.~\ref{fig:alpha_min_sensitivity} varies the minimum alignment weight
$\alpha_{\min}$ on Washington RGB-D. Increasing it from 0 to 0.1
improves performance, with All, Old, and New reaching 88.06\%, 92.26\%,
and 83.87\%, respectively. A zero floor allows low-reliability samples
to receive very small weights, whereas a positive floor preserves
their participation in feature and prediction alignment.

At $\alpha_{\min}=1$, the unlabeled weights become uniform and the
reported All and New accuracies are 81.30\% and 77.18\%. Compared with
0.1, these represent decreases of 6.76 and 6.69 percentage points.
This endpoint removes relative differentiation among unlabeled instance
weights and also changes eligibility for reliability-filtered prototypes.
The observed decrease therefore concerns the combined effect of these
changes, rather than instance weighting alone.

Because feature and prediction losses normalize sample weights within
each batch, the floor changes the relative distribution of supervision,
rather than simply multiplying the total alignment loss. It can also
affect eligibility for reliability-filtered prototype construction.
This distinction separates its role from that of the global RCA weight.
We use $\alpha_{\min}=0.1$ by default.


\paragraph{Overall alignment weight.}

\begin{figure}[t]
    \centering
    \includegraphics[width=0.95\linewidth]
    {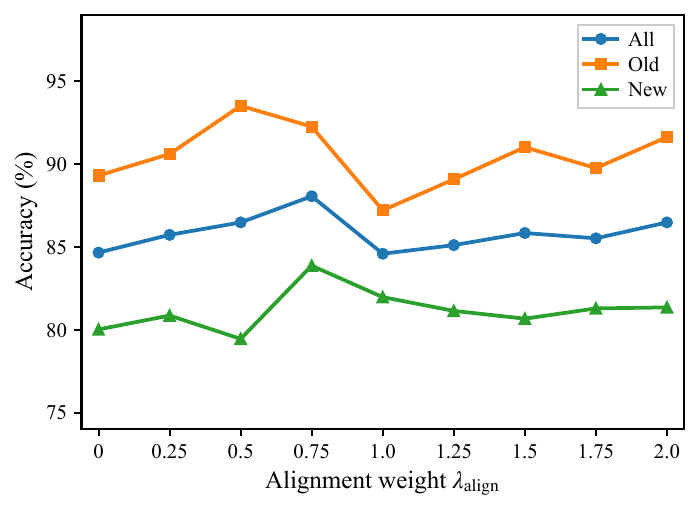}
    \caption{Effect of $\lambda_{\mathrm{align}}$ on Washington RGB-D.}
    \label{fig:lambda_align_sensitivity}
\end{figure}

Fig.~\ref{fig:lambda_align_sensitivity} examines the overall RCA weight
$\lambda_{\mathrm{align}}$ on Washington RGB-D. Setting it to zero
disables RCA and gives 84.66\% All, 89.31\% Old, and 80.02\% New
accuracy. Unlike the sample-weight floor, this coefficient directly
controls RCA's contribution relative to the discovery objective.

The highest All and New values occur at
$\lambda_{\mathrm{align}}=0.75$, reaching 88.06\% and 83.87\%, with
92.26\% Old accuracy. Relative to disabled RCA, the gains are 3.40,
2.95, and 3.85 percentage points on All, Old, and New. Old accuracy
instead peaks at 0.5, where New accuracy is 79.45\%. Choosing the
weight using Old accuracy alone would therefore favor a different
recognition--discovery trade-off.

Larger weights do not consistently improve All or New accuracy.
This pattern supports treating alignment as complementary supervision
whose influence should remain balanced with category discovery.
The reported default is 0.75.

\paragraph{Prototype confidence threshold.}
\label{sec:prototype_threshold}

\begin{figure}[t]
\centering
\includegraphics[width=\columnwidth]{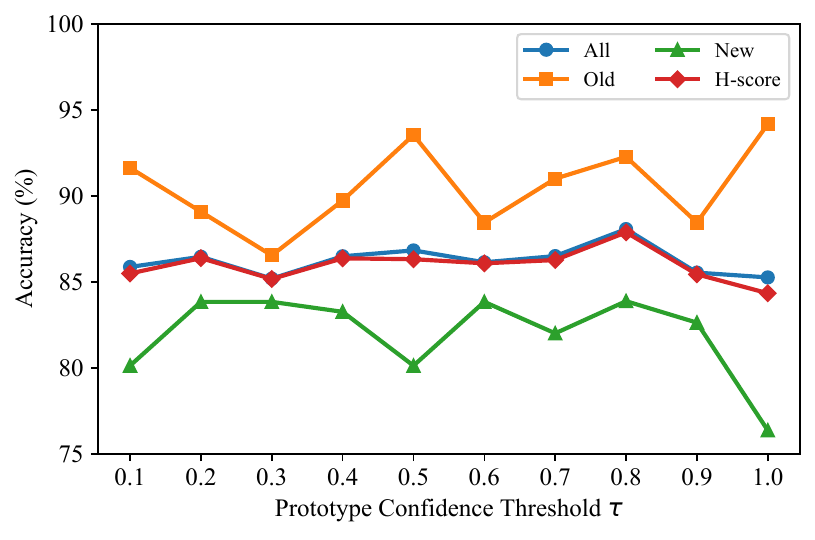}
\caption{Effect of prototype confidence threshold $\tau_p$ on Washington RGB-D.}
\label{fig:proto_threshold_ablation}
\end{figure}

Fig.~\ref{fig:proto_threshold_ablation} examines the prototype confidence
threshold $\tau_p$ on Washington RGB-D. The threshold determines which
fused pseudo-labels are eligible for prototype construction, together
with the separate reliability criterion. Lowering it increases the set
of eligible predictions; raising it excludes less confident assignments.

This trend suggests a trade-off between pseudo-label reliability and
prototype coverage. A permissive threshold can admit incorrect
assignments, while a restrictive threshold can remove useful evidence
and leave fewer pseudo-categories available for transfer.

Among the tested settings, $\tau_p=0.8$ achieves the highest All
accuracy of 88.06\% and an H-score of 87.87\%. The H-score additionally
checks the balance between Old and New rather than evaluating the
threshold solely by overall accuracy. We use $\tau_p=0.8$ by default.

\begin{figure}[t]
    \centering
    \includegraphics[width=\columnwidth]{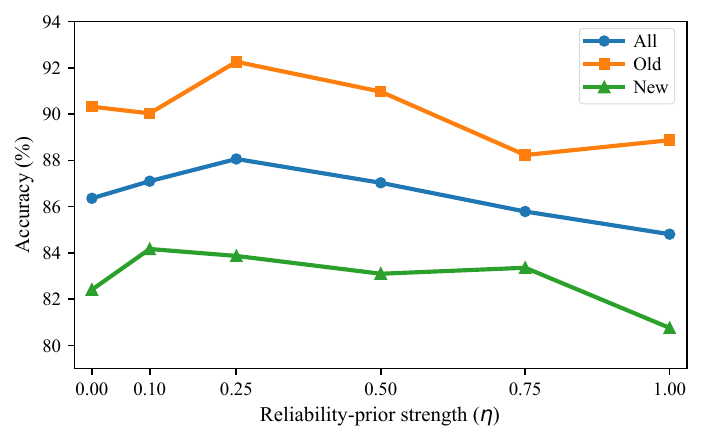}
    \caption{Effect of reliability-prior strength $\eta$ on Washington RGB-D.}
    \label{fig:prior_strength}
\end{figure}

\paragraph{Reliability-prior strength.}
Fig.~\ref{fig:prior_strength} analyzes the influence of the
reliability-prior strength $\eta$ on Washington RGB-D. When the
explicit prior is removed ($\eta=0$), the model achieves 86.36\%,
90.32\%, and 82.41\% accuracy on All, Old, and New categories,
respectively. Introducing a moderate prior improves the overall
performance. In particular, $\eta=0.25$ achieves the best overall
results, with 88.06\%, 92.26\%, and 83.87\% accuracy on All, Old, and
New categories, respectively. Compared with $\eta=0$, this corresponds
to improvements of 1.70, 1.94, and 1.46 points.

A smaller prior strength of $\eta=0.10$ yields the highest New
accuracy of 84.17\%, which is 0.30 points higher than that of the
default setting, but its All and Old accuracy are lower. This suggests
that different prior strengths lead to different trade-offs between
known-category recognition and novel-category discovery, and no single
value maximizes all metrics simultaneously. When $\eta=1.00$, the
allocation is dominated by the explicit prior rather than the learned
channel-wise responses, leading to clear performance degradation,
especially on All and New categories. These results indicate that
moderate prior injection is beneficial, while overly strong reliance
on the handcrafted prior weakens the adaptive allocation mechanism.
Therefore, we set $\eta=0.25$ as the default in all experiments.

\subsection{Input Configuration and Computational Cost}
\label{sec:efficiency}
\label{sec:3.1}

\paragraph{Number of image views.}

Table~\ref{tab:view_number_complexity} varies RGB view count on
Washington RGB-D while fixing the point count at 1,024. From one to
eight views, All accuracy increases from 88.06\% to 93.74\%, and
New accuracy increases from 83.87\% to 91.22\%, giving improvements
of 5.68 and 7.35 percentage points. Additional viewpoints can expose
appearance cues absent from a single observation, providing a plausible
explanation for the discovery gains.

The improvement is not proportional to view count. Moving from four
to eight views adds 0.19 points on All and 0.90 on New, suggesting
diminishing returns in the tested range. Old accuracy peaks at two
views rather than eight, again showing that the input configuration
preferred by known-category recognition need not maximize discovery.

The parameter count remains 179.56 M for all settings, while peak
allocated memory rises from 779.56 to 789.41 MiB. Peak allocation
includes model parameters, input tensors, and intermediate activations
under the reported measurement setup. Measurements use batch-size-one
FP16 inference on an NVIDIA GeForce RTX 4090. The small measured memory
increase characterizes that inference configuration; it does not establish
that training memory or execution time is insensitive to the number of
views. The dagger in the table marks the default single-view setting.


\begin{table}[t]
\centering
\small
\setlength{\tabcolsep}{4.5pt}
\renewcommand{\arraystretch}{1.08}
\caption{Effect of RGB view count on Washington RGB-D.}
\label{tab:view_number_complexity}
\begin{tabular}{ccccccc}
\toprule
\multirow{2}{*}{\textbf{Points}} & \multirow{2}{*}{\textbf{Views}} & \multicolumn{2}{c}{\textbf{Computational Cost}} & \multicolumn{3}{c}{\textbf{Fusion ACC}} \\
\cmidrule(lr){3-4} \cmidrule(lr){5-7}
& & \textbf{Params} & \textbf{Peak Alloc.} & \textbf{All} & \textbf{Old} & \textbf{New} \\
& & \textbf{(M)} & \textbf{(MiB)} & \textbf{(\%)} & \textbf{(\%)} & \textbf{(\%)} \\
\midrule
\multirow{4}{*}{1024}
& 1$^\dagger$ & 179.56 & \textbf{779.56} & 88.06 & 92.26 & 83.87 \\
& 2 & 179.56 & 782.04 & 91.97 & \textbf{98.72} & 85.23 \\
& 4 & 179.56 & 784.54 & 93.55 & 96.77 & 90.32 \\
& 8 & 179.56 & 789.41 & \textbf{93.74} & 96.27 & \textbf{91.22} \\
\bottomrule
\end{tabular}
\vspace{2pt}
\begin{minipage}{0.98\columnwidth}
\footnotesize
$^\dagger$ denotes the default setting.
Memory is measured using batch-size-one FP16 inference on an
NVIDIA GeForce RTX 4090.
\end{minipage}
\end{table}

\paragraph{Number of sampled points.}

Table~\ref{tab:point_number_complexity} varies point count on Washington
RGB-D with a fixed single RGB view. Increasing the count from 256 to
1,024 improves All accuracy from 86.13\% to 88.06\% and New accuracy
from 78.71\% to 83.87\%. The corresponding gains of 1.93 and 5.16
percentage points indicate a larger benefit for discovery in this range.

Increasing the count further to 2,048 matches the All accuracy of
1,024 points at 88.06\%. Old accuracy rises
to 95.48\%, whereas New falls to 80.65\%. The result therefore favors
known-category recognition more than novel-category discovery and does
not support a monotonic benefit from denser point sampling.

The parameter count is unchanged across the tested resolutions.
Peak memory is lower at 256 points and nearly constant from 512 to
2,048 under the same batch-size-one FP16 inference protocol used for
the view-count comparison. This observation describes measured memory,
not the amount of geometric information or the runtime cost.
We retain one RGB view and 1,024 points as the default Washington RGB-D
configuration, marked by a dagger in the table. The other datasets use
their own input settings in Table~\ref{tab:dataset_statistics}.

\begin{table}[t]
\centering
\small
\setlength{\tabcolsep}{3.5pt}
\renewcommand{\arraystretch}{1.08}
\caption{Effect of point count on Washington RGB-D.}
\label{tab:point_number_complexity}
\begin{tabular}{ccccccc}
\toprule
\multirow{2}{*}{\textbf{Views}} & \multirow{2}{*}{\textbf{Points}} & \multicolumn{2}{c}{\textbf{Computational Cost}} & \multicolumn{3}{c}{\textbf{Fusion ACC}} \\
\cmidrule(lr){3-4} \cmidrule(lr){5-7}
& & \textbf{Params} & \textbf{Peak Alloc.} & \textbf{All} & \textbf{Old} & \textbf{New} \\
& & \textbf{(M)} & \textbf{(MiB)} & \textbf{(\%)} & \textbf{(\%)} & \textbf{(\%)} \\
\midrule
\multirow{4}{*}{1}
& 256 & 179.56 & \textbf{739.13} & 86.13 & 93.55 & 78.71 \\
& 512 & 179.56 & 779.55 & 87.38 & 91.61 & 83.17 \\
& 1024$^\dagger$ & 179.56 & 779.56 & \textbf{88.06} & 92.26 & \textbf{83.87} \\
& 2048 & 179.56 & 779.58 & \textbf{88.06} & \textbf{95.48} & 80.65 \\
\bottomrule
\end{tabular}
\vspace{2pt}
\begin{minipage}{0.98\columnwidth}
\footnotesize
$^\dagger$ denotes the default setting.
Memory is measured using batch-size-one FP16 inference on an
NVIDIA GeForce RTX 4090.
\end{minipage}
\end{table}

\subsection{Qualitative Analysis}
\label{sec:qualitative}
\begin{figure*}[!t]
    \centering
    \includegraphics[width=\textwidth]{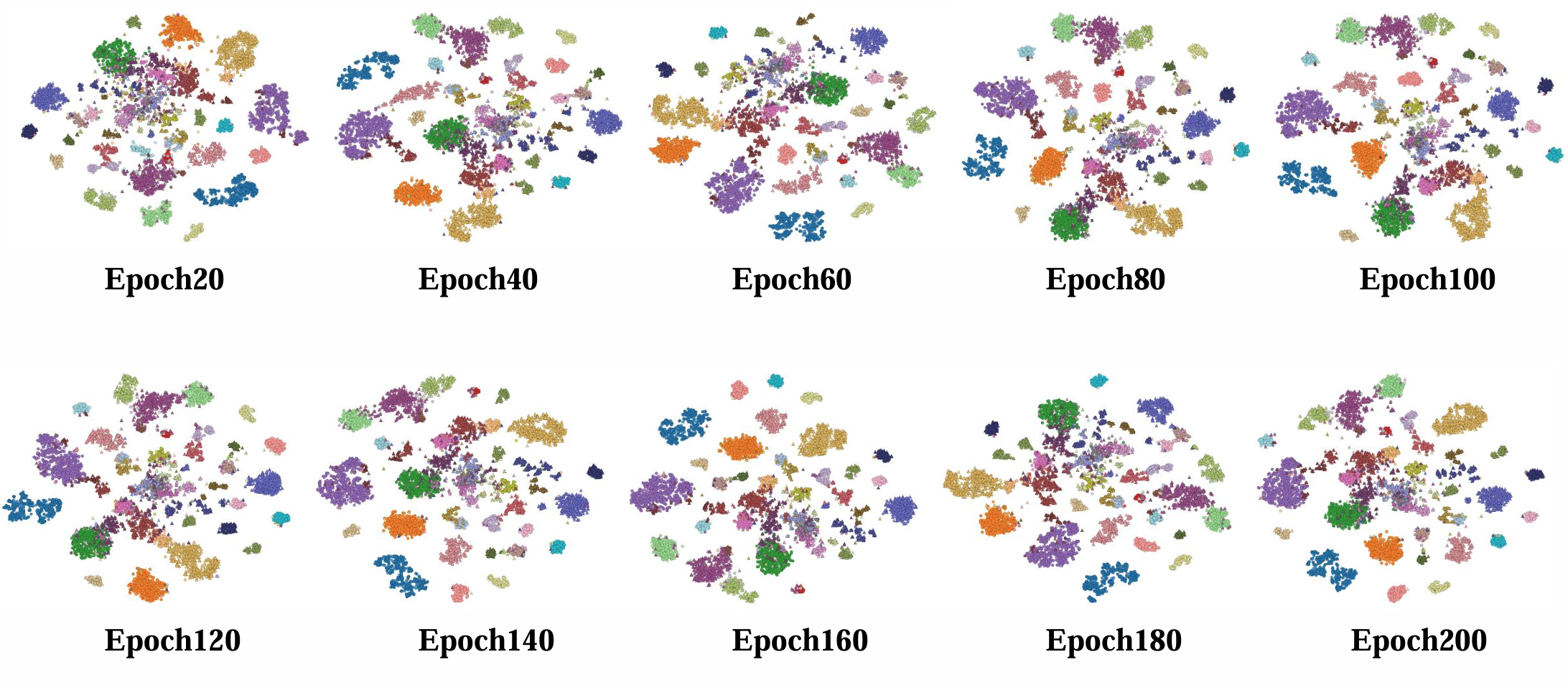}
    \caption{Evolution of fused features on ModelNet40.}
    \label{fig:tsne_fusion}
\end{figure*}

Fig.~\ref{fig:tsne_fusion} shows fused features on ModelNet40 from
epoch 20 to epoch 200. Later checkpoints display more visibly separated
groups, while some mixed regions remain. These plots illustrate the
organization of the learned representation; the quantitative component
ablations establish the performance differences. Since the t-SNE
maps are fitted independently, changes in plot position are not feature
trajectories or quantitative inter-class distances.

\section{Limitations and Future Work}
Although RaGCD demonstrates consistent improvements across diverse 2D--3D benchmarks, several limitations remain. First, the current framework assumes that each image is accurately paired with a corresponding point cloud, whereas real-world multimodal data may contain missing, weakly aligned, or asynchronous observations. Second, following the standard GCD evaluation protocol, the total number of known and novel categories is assumed to be available during clustering, which may restrict its applicability to fully open-world scenarios. Third, the reliability estimates are mainly derived from model predictions and cross-modal consistency. Although effective in our experiments, these signals may become less reliable under severe distribution shifts or when both modalities are simultaneously corrupted. The present experiments evaluate discovery accuracy and allocation responses, but do not directly quantify reliability-score ranking or pseudo-label selection quality. These properties remain to be established through dedicated diagnostics. In addition, maintaining two pretrained encoders and performing feature-, prediction-, and prototype-level alignment introduce additional computational and memory costs. Future work will investigate category-number-free discovery, calibrated or uncertainty-aware reliability estimation, efficient multimodal interaction, and learning with missing or unpaired modalities. Extending RaGCD to scene-level understanding, temporal observations, additional modalities, and continuously evolving open-world environments also represents a promising direction.

\section{Conclusion}
We studied paired 2D--3D generalized category discovery and proposed
RaGCD to incorporate reliability information into both modality
allocation and cross-modal learning. RMA combines feature content and
predictive cues to allocate channel-wise modality contributions. RCA
weights instance-level alignment and selects observations for
prototype transfer. Experiments on three datasets report improvements
in overall and novel-category clustering accuracy, while component
ablations examine the complementary roles of RMA and RCA. Controlled
corruption tests show allocation shifts toward the less corrupted
modality. Prediction-derived reliability remains heuristic, motivating
further study of shared errors, calibration, and discovery with
incomplete modalities.

\bibliographystyle{IEEEtran}
\bibliography{references}
\end{document}